\documentclass[amsmath,amssymb]{revtex4}
\usepackage{graphicx}
\usepackage{bm}
\makeatletter \@addtoreset{equation}{section} \makeatother
\renewcommand{\theequation}{\arabic{section}.\arabic{equation}}

\begin{document}
\title{Dynamics of rogue waves in the Davey-Stewartson II equation}
\author{Yasuhiro Ohta$^{1}$\footnote{Email: ohta@math.kobe-u.ac.jp}
\hspace{0.15cm} and \hspace{0.1cm}  Jianke
Yang$^{2}$\footnote{Email: jyang@math.uvm.edu}} \affiliation{
{\small\it $^1$Department of Mathematics, Kobe University, Rokko,
Kobe 657-8501, Japan} \\ {\small\it $^2$Department of Mathematics
and Statistics, University of Vermont, Burlington, VT $05401$,
U.S.A}}

\begin{abstract}
General rogue waves in the Davey-Stewartson-II equation are derived
by the bilinear method, and the solutions are given through
determinants. It is shown that the simplest (fundamental) rogue
waves are line rogue waves which arise from the constant background
in a line profile and then retreat back to the constant background
again. It is also shown that multi-rogue waves describe the
interaction between several fundamental rogue waves, and
higher-order rogue waves exhibit different dynamics (such as rising
from the constant background but not retreating back to it). A
remarkable feature of these rogue waves is that under certain
parameter conditions, these rogue waves can blow up to infinity in
finite time at isolated spatial points, i.e., exploding rogue waves
exist in the Davey-Stewartson-II equation.

\end{abstract}

\maketitle

\section{Introduction}

Rogue waves are large and spontaneous nonlinear waves and have been
found in a variety of physical systems (such as the ocean and
optical systems) \cite{rogue_water,Rogue_nature1}. Rogue waves
generally occur due to modulation instability of monochromatic
waves. One of the simplest mathematical models for modulation
instability is the nonlinear Schr\"odinger (NLS) equation. For this
equation, explicit expressions of rogue-wave solutions have been
obtained by a variety of techniques such as the Darboux
transformation, the bilinear method and so on
\cite{Peregrine,Akhmediev_PRE,Rogue_higher_order,Rogue_higher_order2,Rogue_Gaillard,Rogue_triplet,Rogue_circular,Liu_qingping,
OY}. These NLS rogue waves can also be obtained from homoclinic
solutions of the NLS equation under certain limits
\cite{Akhmediev_1985,Akhmediev_1988,Its_1988,Ablowitz_homo,Rogue_homo},
or from rational solutions of the Davey-Stewartson equation through
dimension reductions \cite{OY,Ablowitz_private}. Physically these
NLS rogue waves have been observed in optical fibers and water tanks
\cite{Rogue_nature2,NLS_rogue_water}. In addition to the NLS
equation, rogue waves have also been obtained in other wave
equations, such as the Hirota equation, the derivative NLS equation
and the Davey-Stewartson-I equation \cite{rogue_Hirota, rogue_DNLS1,
rogue_DNLS2, OY_DSI}. Explicit rogue-wave solutions in mathematical
model equations reveal the conditions for rogue-wave formation and
facilitate the observation and prediction of rogue waves in physical
systems
\cite{rogue_water,Rogue_nature1,Rogue_nature2,NLS_rogue_water}.

In this article, we derive general rogue-wave solutions in the
Davey-Stewartson-II equation. This equation arises in the modeling
of two-dimensional shallow water waves
\cite{Benney_Roskes,Davey_Stewartson,Ablowitz_book}. Our derivation
uses the bilinear method, and the solutions are expressed in terms
of determinants. We show that the simplest (fundamental) rogue waves
are line rogue waves which arise from the constant background in a
line profile and then retreat back to the constant background again.
We also show that the interaction between several fundamental rogue
waves are described by multi-rogue-wave solutions. However,
higher-order rogue waves are found to exhibit different dynamics,
such as rising from the constant background but not retreating back
to it. An important feature about these rogue waves is that, under
certain parameter conditions, these waves can blow up to infinity in
finite time at isolated spatial points (we call such solutions
exploding rogue waves). The existence of exploding rogue waves is
remarkable, and their appearance can be catastrophic in physical
systems.

It is noted that rogue waves are rational solutions of nonlinear
systems in general. For the Davey-Stewartson equations, certain
types of rational solutions have been derived before \cite{SA}.
Those rational solutions, under parameter restrictions, would yield
multi-rogue waves (see Sec. \ref{sec:muiti} of this article). The
rational solutions we would derive (in the next section), on the
other hand, are more general; and these rational solutions, under
parameter restrictions, could yield not only multi-rogue waves but
also higher-order rogue waves.

\section{Rational solutions in the Davey-Stewartson-II equation}

Evolution of a two-dimensional wavepacket on water of finite depth
is governed by the Benney-Roskes-Davey-Stewartson equation
\cite{Benney_Roskes,Davey_Stewartson,Ablowitz_book}. In the
shallow-water (or long-wave) limit, this equation is integrable (see
\cite{Ablowitz_book2} and the references therein). This integrable
equation is sometimes just called the Davey-Stewartson (DS) equation
in the literature. The DS equation is divided into two types, DSI
and DSII equations, depending on whether the surface tension is
strong or weak \cite{Ablowitz_book}.

In this paper, we study the DSII equation. The normalized form of
this equation is
\begin{equation} \label{DSII}
\begin{array}{l}
iA_t=A_{xx}-A_{yy}+(\epsilon |A|^2-2Q)A,
\\[5pt]
Q_{xx}+Q_{yy}=\epsilon(|A|^2)_{xx},
\end{array}
\end{equation}
where $\epsilon=1$ or $-1$. Through the variable transformation
\begin{equation} \label{vartr}
A=\sqrt{2}\, \frac{g}{f},\quad Q=\epsilon-(2\log f)_{xx},
\end{equation}
where $f$ is a real variable and $g$ a complex one, this equation is
transformed into the bilinear form,
\begin{equation} \label{bilinDSII}
\begin{array}{l}
(D_x^2-D_y^2-iD_t)g\cdot f=0,
\\[5pt]
(D_x^2+D_y^2)f\cdot f=2\epsilon(f^2-|g|^2).
\end{array}
\end{equation}

Rogue waves are rational solutions under certain parameter
restrictions. Thus we first present general rational solutions to
the DSII equation in the following theorem. The proof of this
theorem is given in Appendix A.

\vspace{0.3cm} \noindent{\bf Theorem 1} \ The DSII equation
(\ref{DSII}) admits rational solutions (\ref{vartr}) with $f$ and
$g$ given by $2N\times 2N$ determinants
\begin{equation} \label{formula_rational}
f=\tau_0,
\quad
g=\tau_1,
\end{equation}
where
\begin{equation} \label{e:finaltau}
\tau_n=\left|\begin{matrix} m_{ij}^{(n)} &\widehat{m}_{ij}^{(n)} \cr
\noalign{\vskip5pt} \displaystyle\epsilon\hspace{0.05cm}
\overline{\widehat{m}_{ij}^{(-n)}} &\displaystyle
\overline{m_{ij}^{(-n)}}\end{matrix} \right|,
\end{equation}
\begin{eqnarray}
m_{ij}^{(n)}=
\sum_{k=0}^{n_i}c_{ik}(p_i\partial_{p_i}+\xi'_i+n)^{n_i-k}
\sum_{l=0}^{m_j}d_{jl}(q_j\partial_{q_j}+\eta'_j-n)^{m_j-l}
\frac{1}{p_i+q_j},
\end{eqnarray}
\begin{eqnarray}
\widehat{m}_{ij}^{(n)}=
\sum_{k=0}^{n_i}c_{ik}(p_i\partial_{p_i}+\xi'_i+n)^{n_i-k}
\sum_{l=0}^{m_j}\bar d_{jl}(\bar q_j\partial_{\bar q_j}+\overline{\eta'_j}+n)^{m_j-l}
\frac{1}{p_i\bar q_j+\epsilon},
\end{eqnarray}
\begin{equation} \label{d:xii}
\xi'_i=\frac{p_i-\epsilon/p_i}{2}x+\frac{p_i+\epsilon/p_i}{2}\sqrt{-1} \hspace{0.06cm} y+\frac{p_i^2+1/p_i^2}{\sqrt{-1}}t,
\end{equation}
\begin{equation} \label{d:etaj}
\eta'_j=\frac{q_j-\epsilon/q_j}{2}x+\frac{q_j+\epsilon/q_j}{2}\sqrt{-1} \hspace{0.06cm}y-\frac{q_j^2+1/q_j^2}{\sqrt{-1}}t,
\end{equation}
the overbar `$\ \bar{}\ $' represents complex conjugation, $i, j=1,
\dots, N$, $n_i, m_j$ are arbitrary non-negative integers, and $p_i,
q_j, c_{ik}, d_{jl}$ are arbitrary complex constants.

\vspace{0.2cm} \textbf{Remark 1.} \  By a scaling of $f$ and $g$, we
can normalize $c_{i0} = d_{j0}=1$ without loss of generality, thus
hereafter we set $c_{i0} = d_{j0}=1$.

\vspace{0.2cm} \textbf{Remark 2.} \ For $\epsilon=-1$, $f$ in
(\ref{formula_rational}) is non-negative, i.e., $f\ge0$. A proof is
given in Appendix B. Since $f$ is the denominator of the solutions
$A$ and $Q$, the above rational solutions are nonsingular as long as
$f>0$. But it is also possible that $f$ hits zero and the
corresponding solution blows up to infinity at a certain point of
space-time, which we will see later.

\vspace{0.2cm} \textbf{Remark 3.} Rational solutions in the DS
equations have been derived in \cite{SA} before. The nonsingular
rational solutions for the DSII equation in that paper correspond to
special rational solutions in the above theorem with
$n_1=\cdots=n_N=1$ and $m_1=\cdots=m_N=0$.

\vspace{0.2cm} The simplest rational solution is obtained when
$N=1$, $n_1=1$ and $m_1=0$. In this case,
\[
\tau_n=\left|\begin{matrix} m_{11}^{(n)} & \widehat{m}_{11}^{(n)}
\cr \noalign{\vskip5pt}
\displaystyle\epsilon\hspace{0.05cm}\overline{\widehat{m}_{11}^{(-n)}}
&\displaystyle \overline{m_{11}^{(-n)}}\end{matrix} \right|,
\]
where
\[ m_{11}^{(n)}=\frac{1}{p_1+q_1}
\Big(\xi_1'+n-\frac{p_1}{p_1+q_1}+c_{11}\Big),
\]
\[ \widehat{m}_{11}^{(n)}=\frac{1}{p_1\bar
q_1+\epsilon}\Big(\xi_1'+n-\frac{p_1\bar{q}_1}{p_1\bar
q_1+\epsilon}+c_{11}\Big),
\]
and $\xi_1'$ is defined in (\ref{d:xii}). This solution seems to
have three free complex parameters $p_1$, $q_1$ and $c_{11}$, but
$q_1$ can be absorbed into $c_{11}$ by a reparametrization. Indeed,
by defining
\[
\theta=c_{11}-\frac{p_1}{(|p_1|^2-\epsilon)(|q_1|^2-\epsilon)}\left(
\frac{|p_1\bar q_1+\epsilon|^2}{p_1+q_1}-
\frac{\epsilon \hspace{0.04cm} \bar q_1|p_1+q_1|^2}{p_1\bar q_1+\epsilon}\right),
\]
and denoting $p_1=p$, $\xi_1'+\theta=\xi$, we can show that the
terms in $\tau_n$ which are linear in $\xi+n$ and $\bar\xi-n$
vanish, and this $\tau_n$ reduces to
\begin{equation*}
\tau_n=(\xi+n)(\overline{\xi}-n)+\Delta,
\end{equation*}
\begin{equation*}
\xi=ax+by+\omega t+\theta,  \quad \Delta= \frac{-\epsilon|p|^2}{(|p|^2-\epsilon)^2},
\end{equation*}
\begin{equation*}
a\equiv \frac{p-\epsilon/p}{2}, \quad b\equiv \frac{p+\epsilon/p}{2}i, \quad
\omega\equiv \frac{p^2+1/p^2}{i},
\end{equation*}
up to a constant multiplication, thus this new $\tau_n$ yields the
same solution. The solution from this new $\tau_n$ has only two
independent complex parameters $p$ and $\theta$ now. If we separate
the real and imaginary parts of $a, b, \omega$ and $\theta$ as
\begin{equation*}
a=a_1+ia_2, \hspace{0.2cm} b=b_1+ib_2, \hspace{0.2cm} \omega=\omega_1+i\omega_2, \hspace{0.2cm}
\theta=\theta_1+i\theta_2,
\end{equation*}
then the explicit expressions for this solution are
\begin{equation}  \label{e:Axyt}
A(x,y,t)=\sqrt{2}\left[1-\frac{2i(a_2x+b_2y+\omega_2t+\theta_2)+1}{f}\right].
\end{equation}
\begin{equation}  \label{e:Qxyt}
Q(x,y,t)=\epsilon-(2\log f)_{xx},
\end{equation}
where
\begin{equation*}
f=(a_1x+b_1y+\omega_1t+\theta_1)^2+(a_2x+b_2y+\omega_2t+\theta_2)^2+\Delta.
\end{equation*}

This simplest rational solution is nonsingular when $\epsilon=-1$
(where $\Delta>0$). In this case, the solution exhibits two
distinctly different dynamics depending on the parameter value of
$p$.
\begin{enumerate}
\item If $|p|\ne 1$, then it is easy to see that $b/a$ is not
real, hence $b_1/b_2\ne a_1/a_2$. In this case, along the
$[x(t),y(t)]$ trajectory where
\[a_1x+b_1y=-\omega_1t, \quad a_2x+b_2y=-\omega_2t, \]
solutions $(A, Q)$ are constants. In addition, at any given time,
$(A, Q) \to (\sqrt{2}, \epsilon)$ when $(x,y)$ goes to infinity.
Thus the solution is a two-dimensional lump moving on a constant
background \cite{SA}.

\item If $|p|=1$, then $a, b$ are real
but $\omega$ is imaginary. In this case, the solution depends on
$(x,y)$ through the combination $a_1x+b_1y$ and is thus a line
wave. As $t\to \pm \infty$, this line wave goes to a uniform
constant background (as long as $p^2\ne \pm i$); in the
intermediate times, it rises to a higher amplitude. Thus this
line wave is a line rogue wave which ``appears from nowhere and
disappears with no trace".
\end{enumerate}

When $\epsilon=1$, the rational solution
(\ref{e:Axyt})-(\ref{e:Qxyt}) is singular on a certain elliptic
curve in the $(x,y)$ plane for any time $t$, since $\Delta<0$ now.
For this $\epsilon$, the constant-background solution is
modulationally stable \cite{Tajiri_1999}, thus no rogue waves can be
expected. In view of this, we only consider the case of
$\epsilon=-1$ in the remainder of the paper.

\section{Rogue waves in the Davey-Stewartson-II equation}

As we see from the above analysis, rogue waves would result from the
rational solutions in Theorem 1 for $\epsilon=-1$ under certain
parameter conditions. Specifically, to obtain rogue waves, we need
to require $\epsilon=-1$ and
\begin{equation}\label{e:rogue_condition}
|p_j|=1, \ \mbox{if}\ n_j>0; \quad  |q_j|=1, \  \mbox{if}\ m_j>0; \quad 1\le j\le N.
\end{equation}
In this section, we examine the dynamics of these rogue waves in
detail.

\subsection{Fundamental rogue waves}

Fundamental rogue waves in the DSII equation are obtained when one
takes
\begin{equation}\label{e:rogue_fund}
\epsilon=-1, \  N = 1, \ n_1 = 1, \ m_1=0, \ p_1=e^{i\beta}
\end{equation}
in the rational solution (\ref{formula_rational}), with $\beta$
being a real parameter and $p_1^2\ne \pm i$ (i.e., $\cos2\beta\ne
0$). As we have explained in the previous section, this solution is
equivalent to (\ref{e:Axyt})-(\ref{e:Qxyt}). After a shift of time
and space coordinates, $\theta_1$ and $\theta_2$ can be eliminated.
Then in view of $p=e^{i\beta}$, this fundamental rogue wave becomes
\begin{equation} \label{e:Afund}
\hspace{-1cm}
A(x,y,t)=\sqrt{2}\left(
1-\frac{4-16it\cos2\beta}{1+4(x\cos\beta-y\sin\beta)^2+16t^2\cos^22\beta}\right),
\end{equation}
\begin{equation}  \label{e:Qfund}
\hspace{-1cm}
Q(x,y,t)=-1-16\cos^2\beta \frac{1-4(x\cos\beta-y\sin\beta)^2+16t^2\cos^22\beta}{[1+4(x\cos\beta-y\sin\beta)^2+16t^2\cos^22\beta]^2},
\end{equation}
where $\beta$ is a free real parameter.  This solution describes a
line wave with the line oriented in the ($\sin\beta, \cos\beta$)
direction of the $(x, y)$ plane, and the orientation angle is
$\pi/2-\beta$. The width of this line wave is the same for all
$\beta$ values, i.e., the width is angle-independent. At any given
time, this solution is a constant along the line direction (with
fixed $x\cos\beta-y\sin\beta$) and approaches the constant
background away from the center of the line (with
$x\cos\beta-y\sin\beta\to \pm \infty$). When  $t\to \pm \infty$, the
solution $A$ uniformly approaches the constant background
$\sqrt{2}$; but in the intermediate times, $|A|$ reaches maximum
amplitude $3\sqrt{2}$ (i.e., three times the background amplitude)
at the center ($x\cos\beta-y\sin\beta=0$) of the line wave at time
$t = 0$. The speed at which this line wave climbs to its peak
amplitude is proportional to $|\cos2\beta|$, which is
angle-dependent. This fundamental rogue wave is illustrated in Fig.
1 with $\beta=\pi/6$.

\begin{figure}[h!]
\centerline{\includegraphics[width=0.7\textwidth]{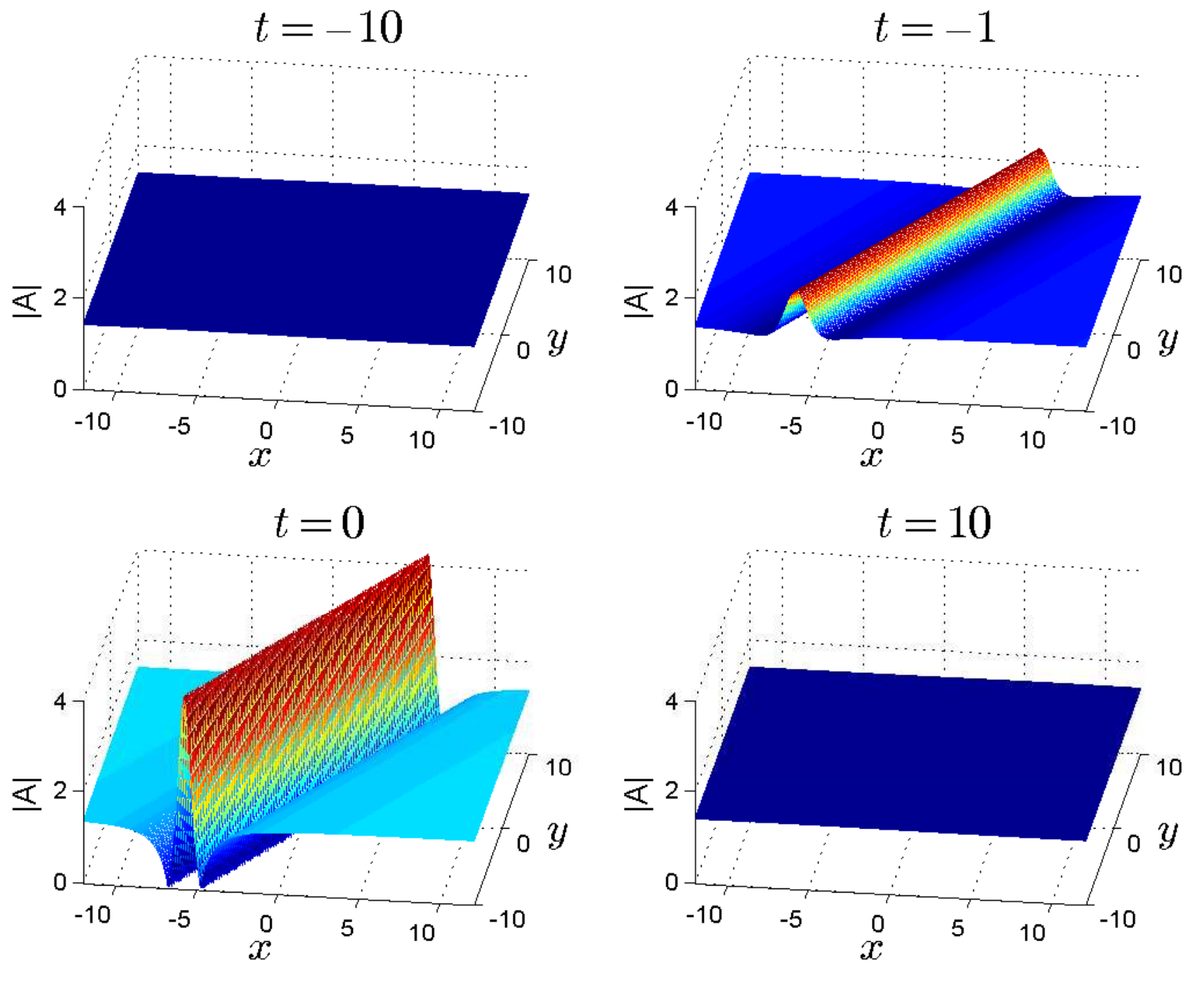}} \caption{A
fundamental rogue wave (\ref{e:Afund}) with $\beta=\pi/6$. }
\label{f:fig1}
\end{figure}

It is noted that under the same parameter conditions
(\ref{e:rogue_fund}) but with $\cos2\beta = 0$, i.e., this line wave
is oriented diagonally ($45^\circ$) or anti-diagonally
($-45^\circ$), then $\omega=0$ in the rational solution
(\ref{e:Axyt})-(\ref{e:Qxyt}). In this case, after a shift of space
coordinates, $\theta_1$ can be eliminated. Hence this rational
solution becomes
\begin{equation} \label{e:AfundS}
\hspace{-1cm}
A(x,y,t)=\sqrt{2}\left(
1-\frac{4+8i\theta_2}{1+2(x\pm y)^2+4\theta_2^2}\right),
\end{equation}
\begin{equation}  \label{e:QfundS}
\hspace{-1cm}
Q(x,y,t)=-1-8 \frac{1-2(x\pm y)^2+4\theta_2^2}{[1+2(x\pm y)^2+4\theta_2^2]^2},
\end{equation}
where $\theta_2$ is a free real parameter. This solution is not a
rogue wave. Instead, it is a stationary line soliton sitting on the
constant background. Its peak $|A|$ amplitude is
$\sqrt{2(9+4\theta_2^2)/(1+4\theta_2^2)}$. The highest value of this
peak amplitude is $3\sqrt{2}$ (three times the constant background),
which is attained at $\theta_2=0$. When $|\theta_2|$ increases to
infinity, this peak amplitude decreases to the background amplitude
$\sqrt{2}$.

If $N>1$, or $N=1$ but $m_1+n_1>1$, the rational solutions in
Theorem 1 under parameter restriction (\ref{e:rogue_condition}) will
give a wide variety of non-fundamental rogue waves. For simplicity,
we consider three subclasses of such solutions below.

\subsection{Multi-rogue waves}  \label{sec:muiti}

One subclass of non-fundamental rogue waves is the multi-rogue waves
which describe the interaction between several fundamental rogue
waves. These solutions can be obtained from Theorem 1 by taking
\begin{equation} \label{cond:multi}
\epsilon=-1, \quad
N> 1, \quad n_j=1, \quad m_j=0, \quad p_j=e^{i\beta_j}, \quad 1\le j\le N,
\end{equation}
where $\beta_j$ is a free real parameter (with $\cos2\beta_j\ne 0$).
In this case, the $\tau$-solution (\ref{e:finaltau}) becomes
\begin{equation} \label{e:finaltau_multi}
\tau_n=\left|\begin{matrix} m_{ij}^{(n)} &\widehat{m}_{ij}^{(n)} \cr
\noalign{\vskip5pt} \displaystyle -\hspace{0.05cm}
\overline{\widehat{m}_{ij}^{(-n)}} &\displaystyle
\overline{m_{ij}^{(-n)}}\end{matrix} \right|,
\end{equation}
where
\begin{equation}
m_{ij}^{(n)}=\frac{1}{p_i+q_j}
\Big(\xi_i'+n-\frac{p_i}{p_i+q_j}+c_{i1}\Big),
\end{equation}
\begin{equation}
\widehat{m}_{ij}^{(n)}=\frac{1}{p_i\bar
q_j-1}\Big(\xi_i'+n-\frac{p_i\bar{q}_j}{p_i\bar
q_j-1}+c_{i1}\Big),
\end{equation}
$\xi_i'$ is defined in (\ref{d:xii}), and $q_j, c_{i1}$ are free
complex constants (but with $q_j\ne \pm p_i$ to avoid zero
divisors). When $t\to \pm \infty$, the solutions $(A, Q)$ approach
the constant background uniformly in the entire $(x, y)$ plane. In
the intermediate times, $N$ fundamental line rogue waves arise from
the constant background, interact with each other, and then
disappear into the background again. Depending on the parameter
choices, individual line rogue waves can reach their peak amplitudes
at the same time or at different times, with the former yielding
stronger interactions.

\begin{figure}[h!]
\centerline{\includegraphics[width=0.7\textwidth]{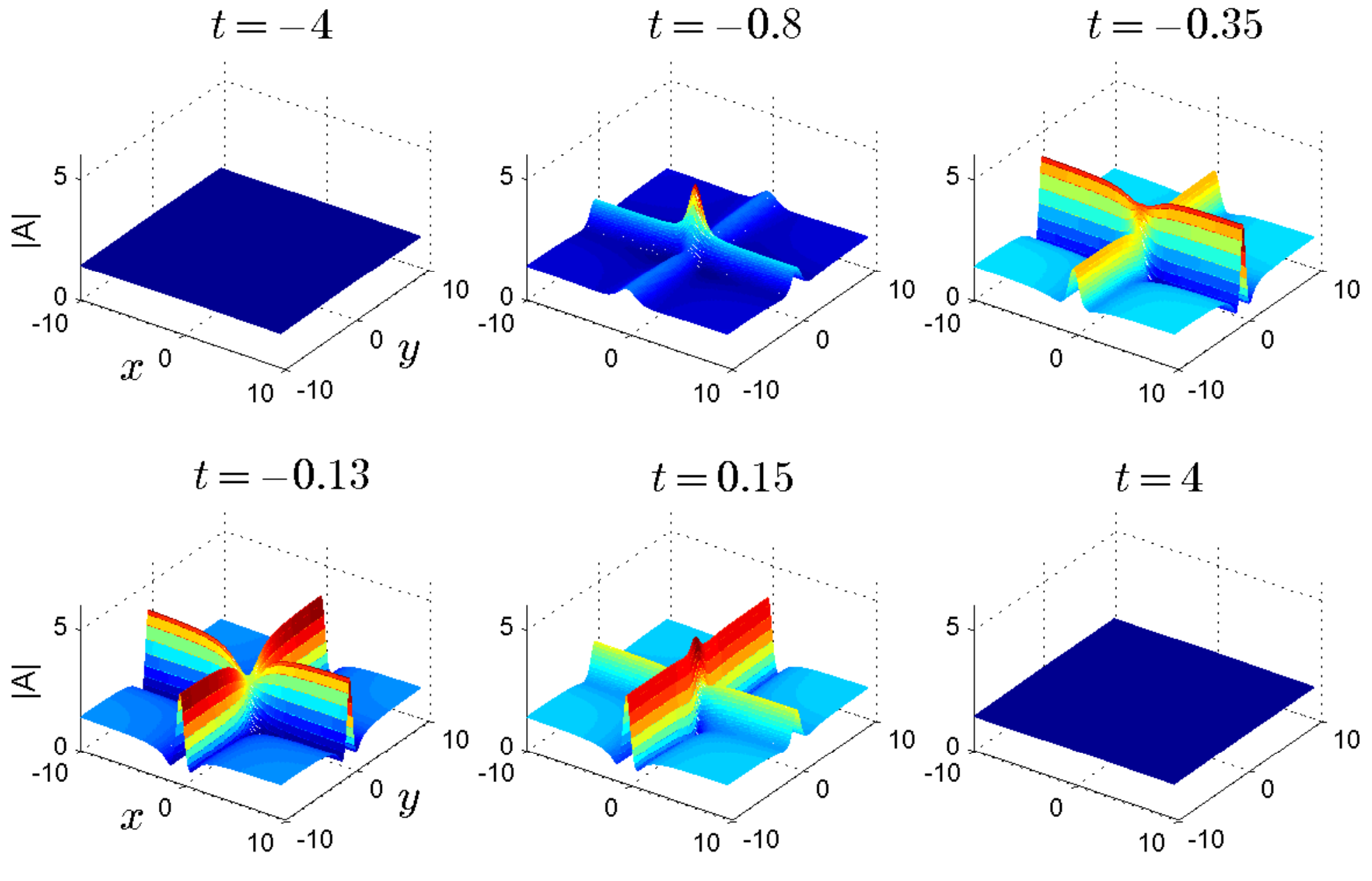}} \caption{A
two-rogue wave solution (\ref{e:finaltau_multi}) with parameters
(\ref{para_fig2}).} \label{f:fig2}
\end{figure}

Now we illustrate these multi-rogue waves and examine their
dynamics. To obtain a two-rogue wave solution, we take parameter
values
\begin{equation} \label{para_fig2}
\hspace{-1cm}
N=2, \quad p_1=1, \quad p_2=i, \quad q_1=0, \quad q_2=-3, \quad c_{11}=0, \quad c_{21}=i/2.
\end{equation}
The corresponding solution $|A|$ is displayed in Fig. \ref{f:fig2}.
It is seen that as $t\to \pm \infty$, the solution uniformly
approaches the constant background $\sqrt{2}$; but in the
intermediate times, a cross-shape rogue wave appears. This cross
rogue wave describes the interaction between two fundamental line
rogue waves, one oriented along the $y$ direction (corresponding to
the parameter $p_1$), and the other one oriented along the $x$
direction (corresponding to the parameter $p_2$). These two
individual line waves reach their peak amplitude $3\sqrt{2}$ at
different times, with the $x$-direction one peaking at $t\approx
-1/4$ and the $y$-direction one peaking at $t\approx 0$.

Next, we take parameter values
\begin{equation} \label{para_fig3a}
N=4, \ p_1=1, \ p_2=e^{i/2}, \ p_3=e^i, \ p_4=e^{2i}, \ q_1=-0.1, \
q_2=0,
\end{equation}
\begin{equation} \label{para_fig3b}
q_3=0.1, \ q_4=0.2,  \  c_{11}=-2i, \ c_{21}=0,  \ c_{31}=2i, \
c_{41}=i/2,
\end{equation}
which gives a four-rogue wave solution. This solution ($|A|$) is
displayed in Fig. 3. As $t\to \pm \infty$, the solution uniformly
goes to the constant background $\sqrt{2}$; but in the intermediate
times, a rogue wave comprising four lines emerges. These four
individual line waves reach their peak amplitudes $3\sqrt{2}$ at
approximately the same time $t=0$, and their widths are identical
(see $t=0$ panel). Due to the interaction of these four line waves,
the maximum amplitude of the solution (at intersections of the four
lines) can be very high. Indeed, at $t=-1$, we find that the peak
amplitude of the solution $|A|$ reaches approximately $30\sqrt{2}$
(i.e., 30 times the constant background). Thus such rogue waves can
be fairly dangerous if they arise in physical situations.

\begin{figure}[h!]
\centerline{\includegraphics[width=0.7\textwidth]{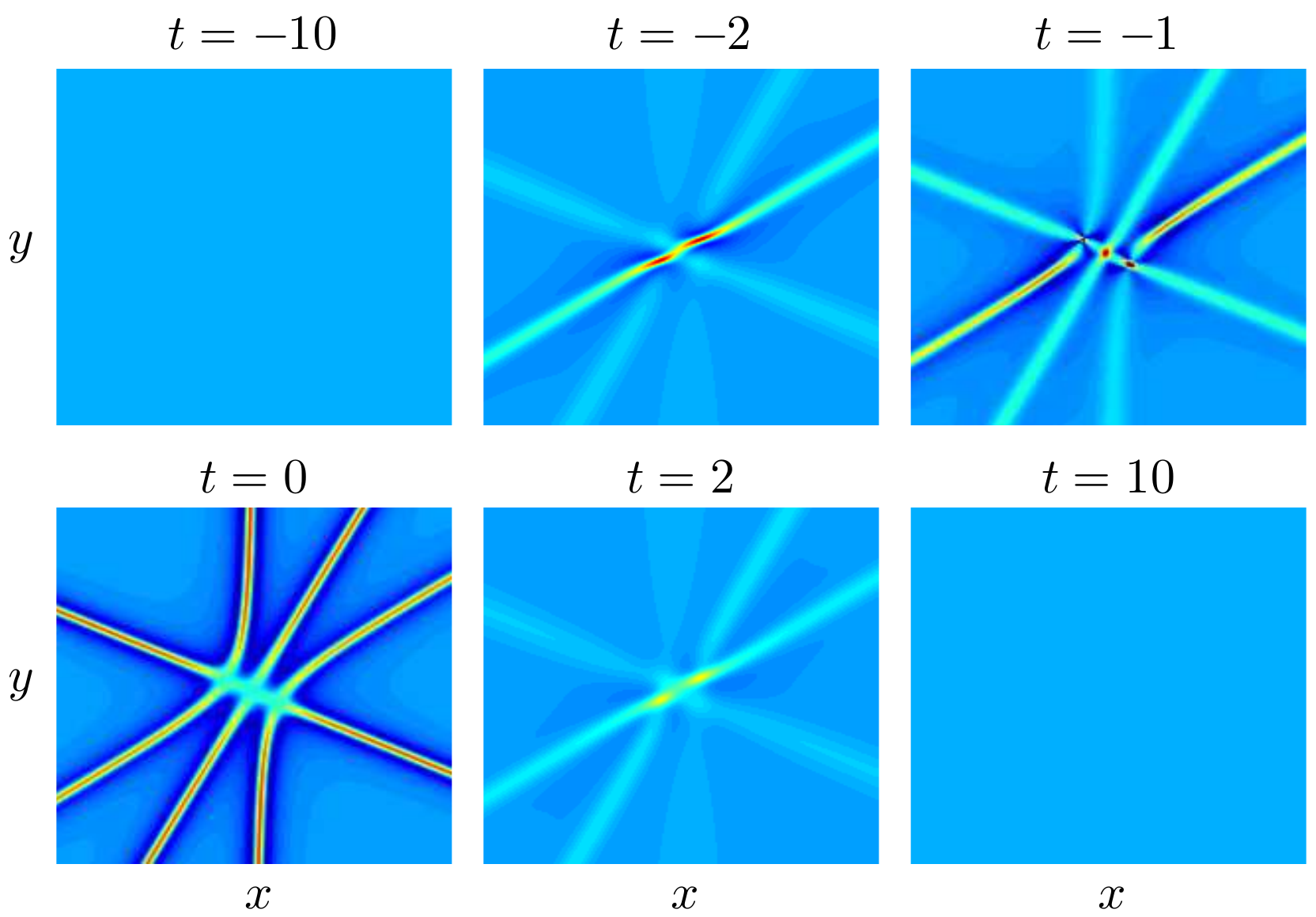}} \caption{A
four-rogue wave solution (\ref{e:finaltau_multi}) with parameters
(\ref{para_fig3a})-(\ref{para_fig3b}). Plotted is the $|A|$ field,
with red color indicating higher values. The spatial region in each
panel is $-20\le x, y\le 20$, and the constant-background value is
$\sqrt{2}$. } \label{f:fig3}
\end{figure}

In the general $N$-rogue wave solution (\ref{e:finaltau_multi}),
$\beta_j$ is a free real parameter, and $q_j, c_{j1}$ are free
complex parameters ($1\le j\le N$). Thus it appears that this
$N$-rogue-wave solution contains $N$ free real parameters and $2N$
free complex parameters, totaling $5N$ free real parameters. But
these parameters are reducible (similar to the simplest rational
solutions in the previous section). Indeed, when $N=2$, by a
reparametrization of
\begin{equation*}
\widehat{c}_{i1}=c_{i1}
-p_i\Big\{\sum_{j=1}^2\Big(\frac{1}{p_i+q_j}+\frac{\bar q_j}{p_i\bar q_j-1}
-\frac{1}{p_i+p_j}\Big)-\frac{1}{p_i-p_{3-i}}\Big\},
\quad i=1,2,
\end{equation*}
we can show that the two-rogue-wave solution
(\ref{e:finaltau_multi}) is reduced to
\begin{equation}   \label{e:reduced_multi}
\hspace{-1cm}
\begin{array}{l}
\displaystyle
\tau_n=\Big((\zeta_1+n)(\zeta_2+n)-\frac{1}{|p_1-p_2|^2}\Big)
\Big((\bar\zeta_1-n)(\bar\zeta_2-n)-\frac{1}{|p_1-p_2|^2}\Big)
\\[5pt]\quad\displaystyle
+\sum_{i=1}^2\sum_{j=1}^2\frac{1}{|p_{3-i}+p_{3-j}|^2}
(\zeta_i+n)(\bar\zeta_j-n)
+\frac{1}{2}\frac{1}{|p_1+p_2|^2}
+\frac{1}{16}\Big|\frac{p_1-p_2}{p_1+p_2}\Big|^4
\end{array}
\end{equation}
up to a constant multiplication (which does not affect the
solution). Here $\zeta_i=\xi'_i+\widehat{c}_{i1},\quad i=1,2$. In
this equivalent $\tau_n$ solution, parameters $q_j$ disappear, thus
it contains only $\beta_1$, $\beta_2$, $\widehat{c}_{11}$ and
$\widehat{c}_{21}$. Of the two complex constants  $\widehat{c}_{11}$
and $\widehat{c}_{21}$, their real parts and the imaginary part of
one of them can be further normalized to be zero by a shift of the
$(x,y,t)$ axes. Thus this two-rogue wave solution contains only
three irreducible real parameters. For the general $N$-rogue wave
solution (\ref{e:finaltau_multi}), we conjecture that all $q_j$
parameters can also be removed by a reparametrization of $c_{i1}$,
hence this $N$-rogue wave solution contains only $3(N-1)$
irreducible real parameters (after a shift of  $(x,y,t)$).

It is noted that if instead of (\ref{cond:multi}),  one takes
\begin{equation*}
N> 1, \quad n_1 =n_2=\dots=n_N=0, \quad m_1=m_2=\dots=m_N=1,
\end{equation*}
then the same multi-rogue-wave solutions as above will be obtained.
Thus different parameter choices can lead to the same solutions.

\subsection{Higher-order rogue waves}

A second subclass of non-fundamental rogue waves is the higher-order
rogue waves. These solutions are obtained from Theorem 1 by taking
\begin{equation}
\epsilon=-1, \quad
N=1, \quad n_1>1, \quad m_1=0, \quad |p_1|=1.
\end{equation}
In this case, the $\tau$-solution (\ref{e:finaltau}) becomes
\begin{equation} \label{e:finaltau_higher}
\tau_n=\left|\begin{matrix} m_{11}^{(n)} &\widehat{m}_{11}^{(n)} \cr
\noalign{\vskip5pt} \displaystyle -\hspace{0.05cm}
\overline{\widehat{m}_{11}^{(-n)}} &\displaystyle
\overline{m_{11}^{(-n)}}\end{matrix} \right|,
\end{equation}
where
\begin{eqnarray}
m_{11}^{(n)}=
\sum_{k=0}^{n_1}c_{1k}(p_1\partial_{p_1}+\xi'_1+n)^{n_1-k}
\frac{1}{p_1+q_1},
\end{eqnarray}
\begin{eqnarray}
\widehat{m}_{11}^{(n)}=
\sum_{k=0}^{n_1}c_{1k}(p_1\partial_{p_1}+\xi'_1+n)^{n_1-k}
\frac{1}{p_1\bar q_1-1},
\end{eqnarray}
$\xi_i'$ is defined in (\ref{d:xii}), $c_{10}=1$, and $c_{1k}, q_1$
are free complex constants. These higher-order rogue waves exhibit
dynamics different from those of multi-rogue waves, as we will
demonstrate below.

For simplicity, we consider second-order rogue waves where $n_1=2$.
In this case, we find that
\begin{equation*}
\hspace{-1.5cm}
m_{11}^{(n)}=\frac{1}{p_1+q_1}
\Big\{\Big(\xi_1'+n-\frac{p_1}{p_1+q_1}+\frac{c_{11}}{2}\Big)^2
+\xi_1''+c_{12}-\frac{c_{11}^2}{4}
-\frac{p_1q_1}{(p_1+q_1)^2}\Big\},
\end{equation*}
%
%
\begin{equation*}
\hspace{-1.5cm}
\widehat{m}_{11}^{(n)}=\frac{1}{p_1\bar{q}_1-1}
\Big\{\Big(\xi_1'+n-\frac{p_1\bar q_1}{p_1\bar q_1-1}+\frac{c_{11}}{2}\Big)^2
+\xi_1''+c_{12}-\frac{c_{11}^2}{4}
+\frac{p_1\bar q_1}{(p_1\bar q_1-1)^2}\Big\},
\end{equation*}
where
\[
\xi_1''\equiv p_1\partial_{p_1}\xi_1'=
\frac{p_1-1/p_1}{2}x+\frac{p_1+1/p_1}{2}\sqrt{-1} \hspace{0.06cm}
y+\frac{p_1^2-1/p_1^2}{\sqrt{-1}}\hspace{0.05cm} 2t.
\]
Denoting
\[p=p_1, \quad q= q_1, \quad \xi= \xi_1'+a, \quad
\zeta=\xi_1''+b,
\]
where $a\equiv c_{11}/2-1$ and $b\equiv c_{12}-c_{11}^2/4$, the
$\tau_n$ solution (\ref{e:finaltau_higher}) becomes
\begin{eqnarray*}
&& \hspace{-2.5cm} \tau_n=\frac{1}{|p+q|^2}
\Big\{\Big(\xi+n+\frac{q}{p+q}\Big)^2+\zeta-\frac{pq}{(p+q)^2}\Big\}
\Big\{\Big(\bar\xi-n+\frac{\bar q}{\bar p+\bar q}\Big)^2
+\bar\zeta-\frac{\bar p\bar q}{(\bar p+\bar q)^2}\Big\}
\\
&& \hspace{-3.2cm} \qquad
+\frac{1}{|p\bar q-1|^2}
\Big\{\Big(\xi+n-\frac{1}{p\bar q-1}\Big)^2
+\zeta+\frac{p\bar q}{(p\bar q-1)^2}\Big\}
\Big\{\Big(\bar\xi-n-\frac{1}{\bar pq-1}\Big)^2
+\bar\zeta+\frac{\bar pq}{(\bar pq-1)^2}\Big\}.
\end{eqnarray*}
This solution has four apparent complex parameters, $p, q, a$ and
$b$. But $q$ can be removed by a reparametrization of $a$ and $b$.
Indeed, by replacing
\begin{eqnarray*}
&&a\to a-1+p\Big(\frac{1}{p+q}+\frac{\bar q}{p\bar q-1}-\frac{\bar p}{|p|^2+1}\Big),
\\
&&b\to b+p\Big(\frac{q}{(p+q)^2}-\frac{\bar q}{(p\bar q-1)^2}
-\frac{\bar p}{(|p|^2+1)^2}\Big),
\end{eqnarray*}
and recalling $|p|=1$,  the above $\tau_n$ can be rewritten as
\begin{equation} \label{tau:second_order}
\tau_n=\Big((\xi+n)^2+\zeta\Big)\Big((\bar\xi-n)^2+\bar\zeta\Big)
+(\xi+n)(\bar\xi-n)
\end{equation}
up to a constant multiplication. Thus this second-order solution
contains only parameters $p$, $a$ and $b$ now.

In these second-order solutions, if $p^2\ne \pm i$, then the
solutions do not uniformly approach the constant background as $t\to
\pm\infty$, thus they are not rogue waves. But when $p^2=-i$, the
solution uniformly approaches the constant background as $t\to
-\infty$, thus it ``appears from nowhere" and is a rogue wave.
However, this second-order rogue wave does not retreat back to the
constant background when $t\to +\infty$, thus it does \emph{not}
``disappear with no trace". This means that this second-order rogue
wave behaves quite differently from the multi-rogue waves considered
in the previous subsection.

\begin{figure}[h!]
\centerline{\includegraphics[width=0.7\textwidth]{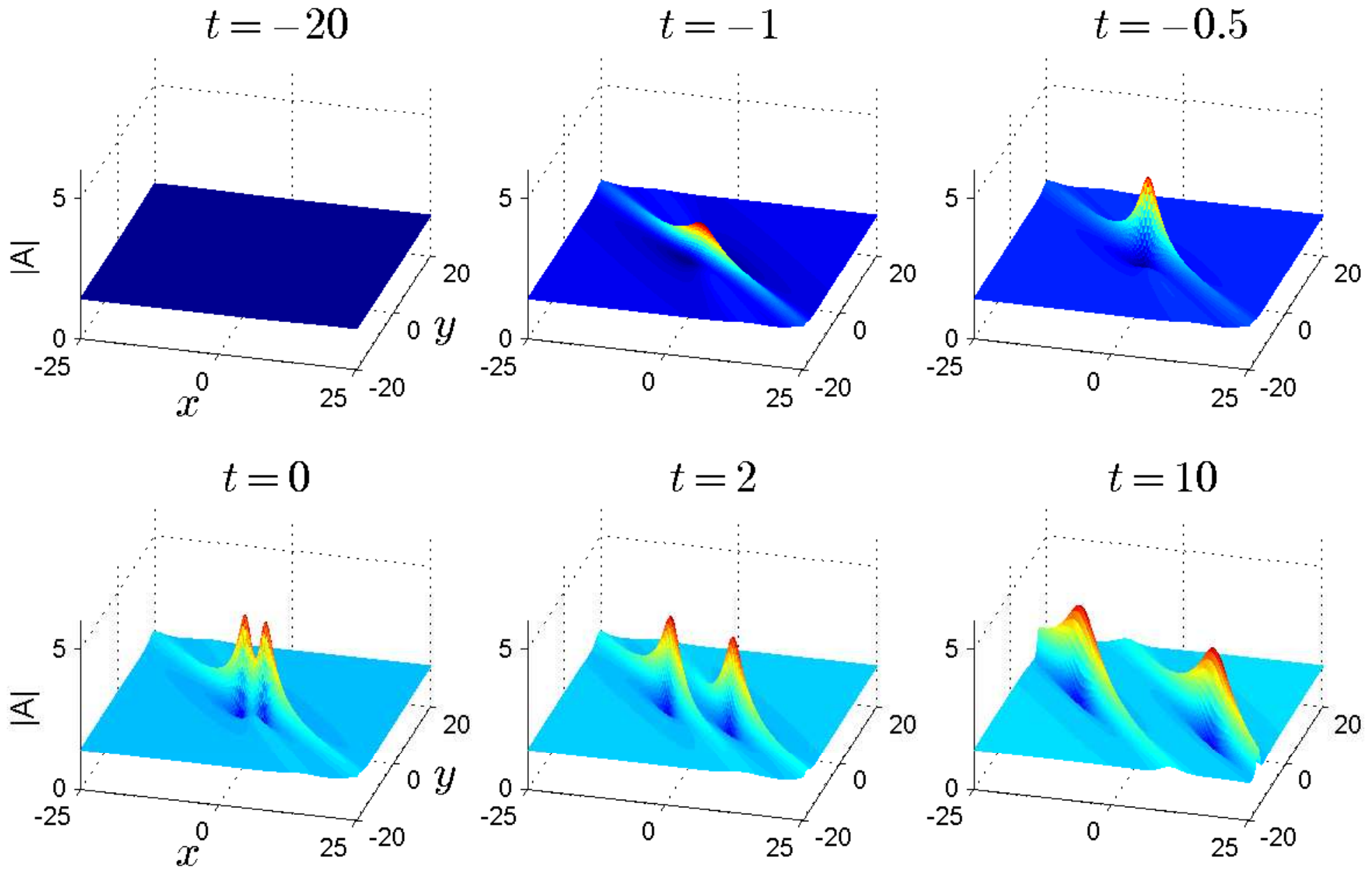}} \caption{A
second-order rogue wave solution (\ref{e:2ndorder}) with
$\alpha=1$.} \label{f:fig4}
\end{figure}

Below we examine this second-order rogue wave in more detail. For
definiteness, we take $p=e^{-i\pi/4}$ (the choice of
$p=-e^{-i\pi/4}$ would yield the same solution). In this case, by a
shift of $(x,y,t)$ axes, we can normalize $b$ as well as the real
part of $a$ to be zero. Thus we can set
\begin{equation}
a=i\alpha, \quad b=0,
\end{equation}
where $\alpha$ is a free real parameter. Substituting these $p$, $a$
and $b$ values into the $\tau_n$ solution (\ref{tau:second_order}),
we find that the solution $A(x,y,t)$ becomes
\begin{equation} \label{e:2ndorder}
\hspace{-2.5cm}
A=\sqrt{2}\left[ 1-\frac{
(1+2i\alpha)[(x+y)^2+8t]-2i(x^2-y^2+\alpha-2\alpha^3)+6\alpha^2} {
\Big(\frac{1}{2}(x+y)^2-4t-\alpha^2\Big)^2+2\Big(\alpha(x+y)-\frac{1}{2}(x-y)\Big)^2+\frac{1}{2}(x+y)^2+\alpha^2}\right],
\end{equation}
and the solution $Q(x,y,t)$ is given by (\ref{vartr}) with
$\epsilon=-1$ and $f$ being the denominator in the above $A$
solution. When $t\to -\infty$, this solution $A(x,y,t)$ uniformly
approaches the constant background $\sqrt{2}$ (like regular rogue
waves). But when $t\to +\infty$, it approaches two lumps which
slowly move away from each other. The peak amplitudes of these two
lumps are attained at $(x,y)$ locations where the first two terms in
the denominator of (\ref{e:2ndorder}) vanish, i.e., at
\[
x_{\mbox{max}}=\pm \Big(\frac{1}{2}+\alpha\Big) \sqrt{8t+2\alpha^2}, \quad
y_{\mbox{max}}=\pm \Big(\frac{1}{2}-\alpha\Big) \sqrt{8t+2\alpha^2},
\]
and these peak $|A|$ amplitudes approach $3\sqrt{2}$ when $t\to
+\infty$. This solution with $\alpha=1$ is displayed in Fig.
\ref{f:fig4}. We see that this second-order rogue wave looks quite
different from the previous rogue waves in Figs. 1-3. Instead of
``disappearing with no trace", this second-order rogue wave
``disappears with a trace".

It is noted that when $p^2=i$, i.e., $p=\pm e^{i\pi/4}$, the
second-order solution (\ref{tau:second_order}) would approach the
constant background when $t\to +\infty$ but approach two lumps which
move away from each other when $t\to -\infty$. In other words, this
solution describes a process which is opposite of that when $p^2=-i$
(see Fig. \ref{f:fig4}).

\subsection{Exploding rogue waves}

A third but important subclass of non-fundamental rogue waves is the
exploding rogue waves. These rogue waves, which arise from the
constant background, can blow up to infinity in finite time at
isolated spatial locations. These exploding rogue waves can be
obtained from the higher-order rogue waves or multi-rogue waves
under certain parameter conditions, as we will demonstrate below.

\begin{figure}[h!]
\centerline{\includegraphics[width=0.7\textwidth]{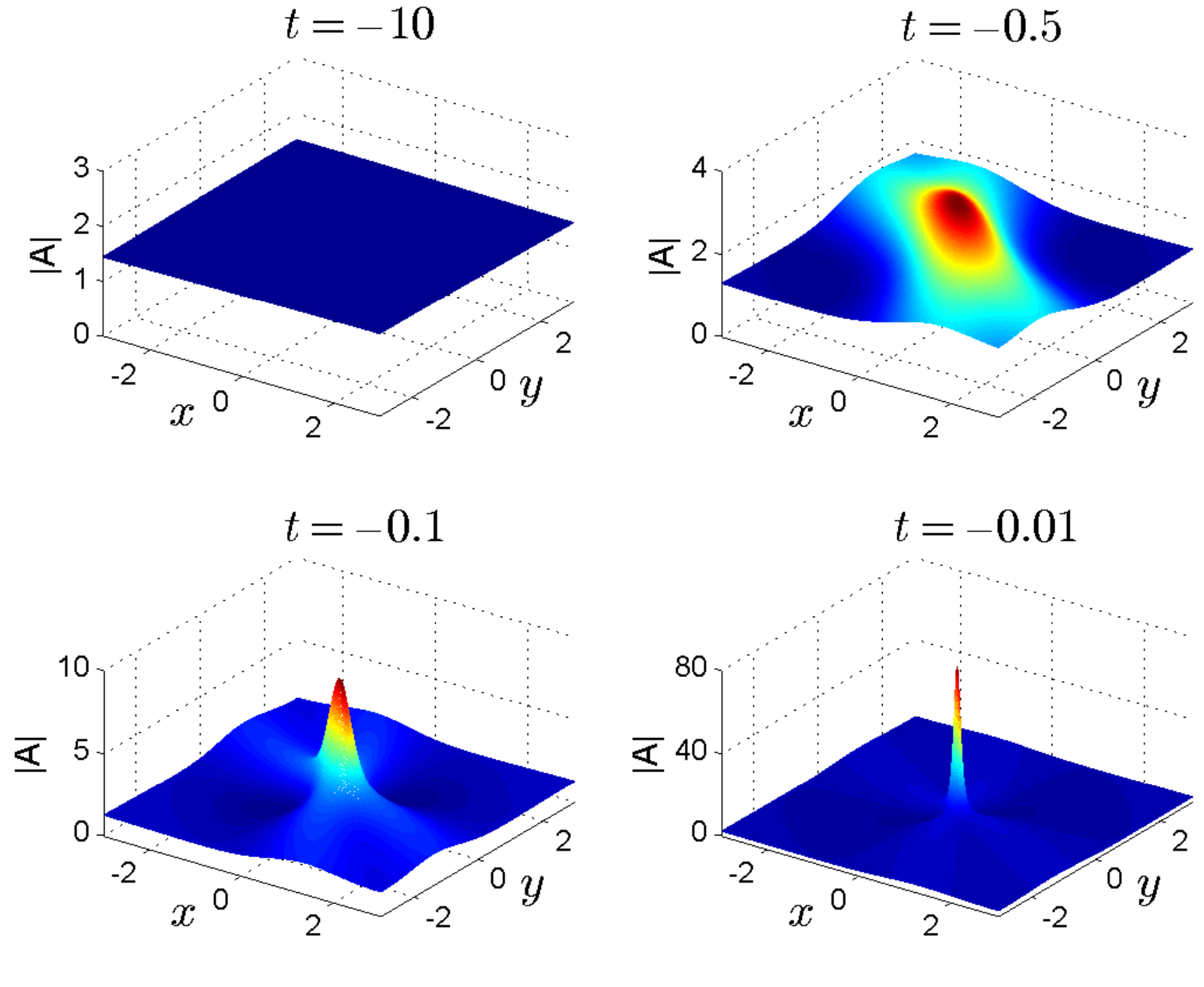}} \caption{An
exploding second-order rogue wave (\ref{e:blowup}). } \label{f:fig5}
\end{figure}

First, we consider the second-order rogue waves (\ref{e:2ndorder}).
When $\alpha=0$, this solution becomes
\begin{equation} \label{e:blowup}
A(x,y,t)=\sqrt{2}\left[ 1-\frac{
(x+y)^2+8t-2i(x^2-y^2)} {
\Big(\frac{1}{2}(x+y)^2-4t\Big)^2+x^2+y^2}\right].
\end{equation}
This solution uniformly approaches the constant background
$\sqrt{2}$ as $t\to \pm \infty$. But in the intermediate time $t=0$,
it blows up to infinity at the origin $(x,y)=(0,0)$. To see this, we
notice that at $(x,y)=(0,0)$,
\begin{equation} \label{e:blowupA00}
A(0,0,t)=\sqrt{2}\hspace{0.04cm} \Big(1-\frac{1}{2t}\Big),
\end{equation}
thus this $A$ solution blows up to infinity when $t$ approaches zero
(the solution $Q$ blows up to infinity at this time as well). The
rate of blowup is $(t-t_*)^{-1}$, where $t_*=0$ is the time of
singularity. This exploding process is displayed in
Fig.~\ref{f:fig5}. The existence of exploding rogue waves in the
DSII equation is a remarkable phenomenon, and their occurrence would
be catastrophic in physical systems.

\begin{figure}[h!]
\centerline{\includegraphics[width=0.7\textwidth]{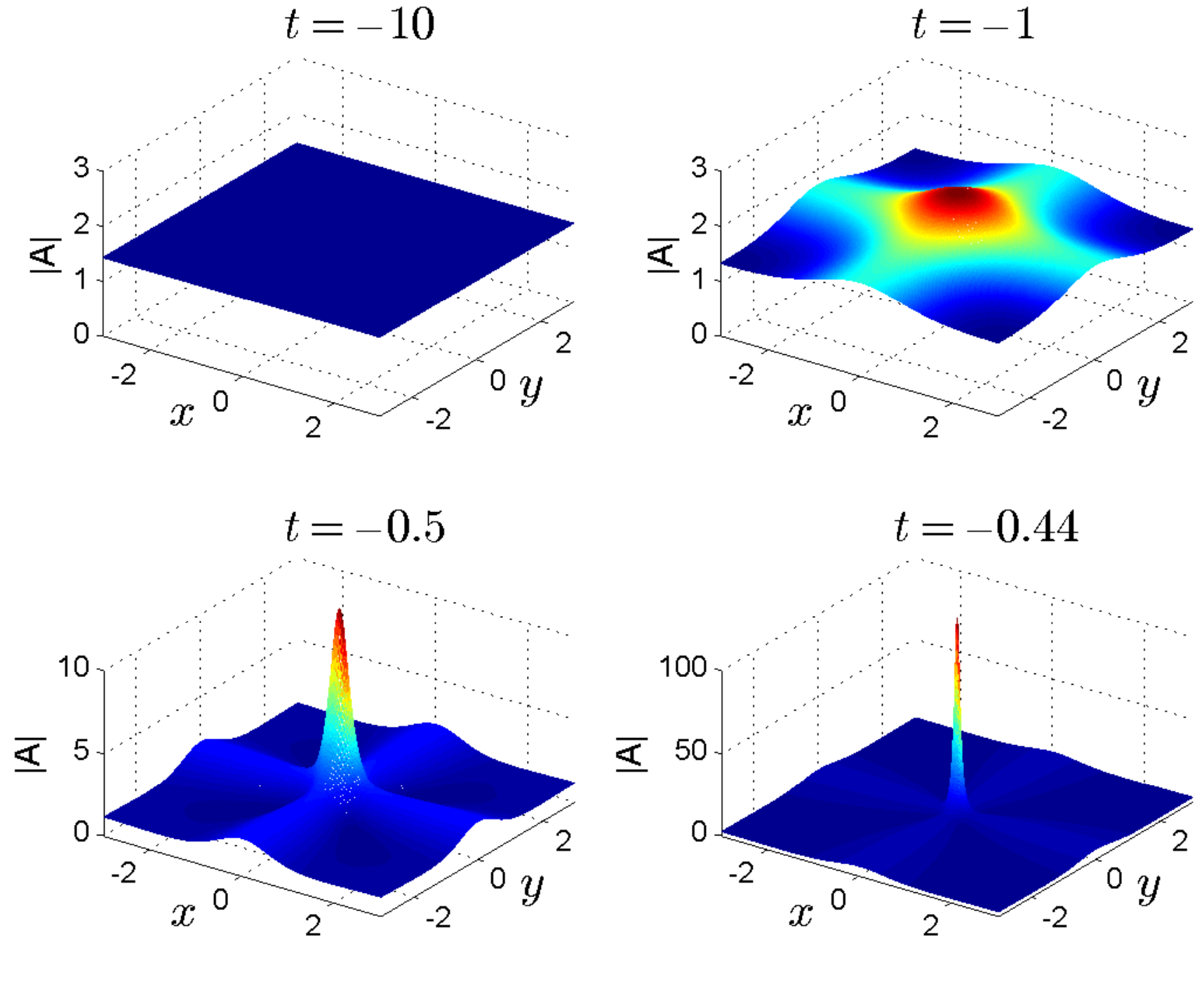}} \caption{An
exploding two-rogue wave (\ref{e:blowup_two_rogue}). }
\label{f:fig6}
\end{figure}

In addition to higher-order rogue waves, multi-rogue waves can also
explode under suitable choices of parameters. To demonstrate, we
consider the two-rogue-wave solutions whose simplified expressions
are given in Eq. (\ref{e:reduced_multi}). Taking parameter values
\begin{equation}
p_1=1, \quad p_2=i, \quad \hat{c}_{11}=\hat{c}_{21}=0,
\end{equation}
this two-rogue wave becomes
\begin{equation} \label{e:blowup_two_rogue}
A(x,y,t)=\sqrt{2} \ \frac{\tau_1}{\tau_0}, \qquad Q=-1-(2\log \tau_0)_{xx},
\end{equation}
where
\[
\tau_0=x^2y^2+\left(4t^2+\frac{1}{4}\right)(x^2+y^2)+\left(4t^2-\frac{3}{4}\right)^2,
\]
\[
\tau_1=x^2y^2+\left(4t^2-\frac{3}{4}\right)\left(x^2+y^2+4t^2+\frac{5}{4}\right)-4it(x^2-y^2).
\]
At the origin $(x,y)=(0,0)$,
\begin{equation}
A(0,0,t)=\sqrt{2} \ \frac{t^2+\frac{5}{16}}{t^2-\frac{3}{16}},
\end{equation}
thus this wave explodes to infinity at times $t_*=\pm \sqrt{3}/4$.
Its exploding rate is also $(t-t_*)^{-1}$, where $t_*$ is the time
of wave singularity. This exploding two-rogue-wave solution is
displayed in Fig. \ref{f:fig6}.

It is noted that for the Davey-Stewartson equations, self-similar
collapsing solutions have been derived in
\cite{Nakamura_1982,Nakamura_1983}. For the non-integrable
Benney-Roskes-Davey-Stewartson equations, wave collapse has also
been reported \cite{Ablowitz_Segur_1979,Ablowitz_collapse2}. Those
collapsing solutions are different from our exploding rogue waves
since the boundary conditions of those solutions are different.
Specifically, those collapsing solutions do not arise from the
constant background and are thus not rogue waves.

In this section, only a few subclasses of rogue-wave solutions were
examined. The rational solutions in Theorem 1, under parameter
conditions (\ref{e:rogue_condition}), also contain a lot of other
subclasses of rogue waves which are not elaborated in this article.
We also note that different choices of parameters can yield the same
solutions. For instance, if we take
\[ N=1, \quad n_1=m_1=1, \quad |p_1|=|q_1|=1 \]
in Theorem 1, the resulting solution (\ref{formula_rational}) would
be equivalent to the two-rogue-wave solution
(\ref{e:finaltau_multi}) with parameters
\[ N=2, \quad n_1=n_2=1, \quad m_1=m_2=0, \quad |p_1|=|p_2|=1 \]
(see also Eq. (\ref{e:reduced_multi})).

\section{Summary and discussions}

In this article, we have derived general rogue waves in the
Davey-Stewartson-II equation. We have shown that the fundamental
rogue waves are line rogue waves which arise from the constant
background in a line profile and then retreat back to the constant
background again. We have also shown that multi-rogue waves describe
the interaction between several fundamental rogue waves, and
higher-order rogue waves exhibit different dynamics (such as rising
from the constant background but not retreating back to it). In
addition, we have discovered exploding rogue waves, which arise from
the constant background but blow up to infinity in finite time at
isolated spatial points.

It is helpful to compare these rogue waves in the
Davey-Stewartson-II equation with those in the Davey-Stewartson-I
equation (see \cite{OY_DSI}). The biggest difference is that
exploding rogue waves exist in the Davey-Stewartson-II equation, but
such waves cannot be found in the Davey-Stewartson-I equation
\cite{OY_DSI}. In Appendix C, nonsingularity of rogue waves in the
Davey-Stewartson-I equation is analytically proved for a subclass of
parameter values, and we conjecture that all rogue waves (which
arise from the constant background) are nonsingular in the
Davey-Stewartson-I equation.

Other differences on rogue waves also exist between the
Davey-Stewartson-I and Davey-Stewartson-II equations. For instance,
in the Davey-Stewartson-I equation, fundamental (line) rogue waves
can only be oriented along a half of all possible angles in the
$(x,y)$ plane \cite{OY_DSI}; but in the Davey-Stewartson-II
equation, fundamental rogue waves can be oriented along any angle
(except diagonal and anti-diagonal angles). This difference has
important implications for multi-rogue-wave patterns. For instance,
in the Davey-Stewartson-II equation, cross rogue-wave patterns
formed by two orthogonally-oriented fundamental rogue waves exist
(see Fig. \ref{f:fig2}); but in the Davey-Stewartson-I equation,
cross patterns of multi-rogue waves cannot exist.

\section*{Acknowledgment}
The work of Y.O. is supported in part by JSPS Grant-in-Aid for
Scientific Research (B-24340029, S-24224001) and for Challenging
Exploratory Research (22656026), and the work of J.Y. is supported
in part by the Air Force Office of Scientific Research (Grant USAF
9550-12-1-0244).

\section*{Appendix A}

\renewcommand{\theequation}{A.\arabic{equation}}
\setcounter{equation}{0}

In this appendix, we derive the rational solutions to the DSII
equation in Theorem 1.

The bilinear form (\ref{bilinDSII}) of the DSII equation can be
derived from
\begin{equation} \label{bilin}
\begin{array}{l}
(D_{x_1}D_{x_{-1}}-2) \tau_n\cdot\tau_n=-2 \tau_{n+1}\tau_{n-1},
\\[5pt]
(D_{x_1}^2-D_{x_2})\tau_{n+1}\cdot\tau_n=0,
\\[5pt]
(D_{x_{-1}}^2+D_{x_{-2}})\tau_{n+1}\cdot\tau_n=0,
\end{array}
\end{equation}
by taking the independent and dependent variables as
\begin{equation} \label{indepvar}
x_1=\frac{1}{2}(x+iy),\quad x_{-1}=\frac{\epsilon}{2}(x-iy),\quad
x_{2}=\frac{1}{2i}t,\quad x_{-2}=-\frac{1}{2i}t,
\end{equation}
\begin{equation} \label{depvar}
f=\tau_0,\quad g=\tau_1,
\end{equation}
and imposing the complex conjugate condition
\begin{equation} \label{cc}
\overline{\tau_n}=\tau_{-n}.
\end{equation}
The variable transformation (\ref{indepvar}) means that
\begin{equation} \label{relx}
x_{-1}=\epsilon\overline{x_1},\quad x_{-2}=\overline{x_2}.
\end{equation}
We first consider the rational solutions for the system
(\ref{bilin}), and then obtain those for (\ref{bilinDSII}) by
imposing the complex conjugate condition (\ref{cc}).

It is known that the bilinear equation (\ref{bilin}) admits
determinant solutions
\begin{equation} \label{e:taunK}
\tau_n=\det_{1\le i,j\le K}\left(m_{ij}^{(n)}\right),
\end{equation}
where $K$ is a positive integer, $m_{ij}^{(n)}$ is an arbitrary
function satisfying the differential and difference relations,
\begin{equation} \label{relm}
\begin{array}{l}
\partial_{x_1}m_{ij}^{(n)}=\varphi_i^{(n)}\psi_j^{(n)},
\\[5pt]
\partial_{x_2}m_{ij}^{(n)}
=\varphi_i^{(n+1)}\psi_j^{(n)}+\varphi_i^{(n)}\psi_j^{(n-1)},
\\[5pt]
\partial_{x_{-1}}m_{ij}^{(n)}=-\varphi_i^{(n-1)}\psi_j^{(n+1)},
\\[5pt]
\partial_{x_{-2}}m_{ij}^{(n)}
=-\varphi_i^{(n-2)}\psi_j^{(n+1)}-\varphi_i^{(n-1)}\psi_j^{(n+2)},
\\[5pt]
m_{ij}^{(n+1)}=m_{ij}^{(n)}+\varphi_i^{(n)}\psi_j^{(n+1)},
\end{array}
\end{equation}
and $\varphi_i^{(n)}$, $\psi_j^{(n)}$ are arbitrary functions
satisfying
\begin{equation} \label{relphipsi}
\partial_{x_{\nu}}\varphi_i^{(n)}=\varphi_i^{(n+\nu)},\quad
\partial_{x_{\nu}}\psi_j^{(n)}=-\psi_j^{(n-\nu)},\quad (\nu=\pm 1,\pm 2).
\end{equation}
The rational solutions are obtained by taking
\begin{equation} \label{mphipsi}
m_{ij}^{(n)}=A_iB_j\frac{1}
{p_i+q_j}\Big(-\frac{p_i}{q_j}\Big)^ne^{\xi_i+\eta_j},
\end{equation}
\begin{equation}
\varphi_i^{(n)}=A_ip_i^ne^{\xi_i},\quad\psi_j^{(n)}=B_j(-q_j)^{-n}e^{\eta_j},
\end{equation}
\begin{equation}
\xi_i=\frac{1}{p_i^2}x_{-2}+\frac{1}{p_i}x_{-1}+p_ix_1+p_i^2x_2,
\end{equation}
\begin{equation}
\eta_j=-\frac{1}{q_j^2}x_{-2}+\frac{1}{q_j}x_{-1}+q_jx_1-q_j^2x_2,
\end{equation}
where $p_i$ and $q_j$ are complex constants, $A_i$ and $B_j$ are
differential operators of order $n_i$ and $m_j$ with respect to
$p_i$ and $q_j$ respectively, defined as
\begin{equation} \label{AB}
A_i=\sum_{k=0}^{n_i}c_{ik}(p_i\partial_{p_i})^{n_i-k},\quad
B_j=\sum_{l=0}^{m_j}d_{jl}(q_j\partial_{q_j})^{m_j-l},
\end{equation}
$c_{ik}$, $d_{jl}$ are complex constants, and $n_i$, $m_j$ are
non-negative integers. It is easy to see that the above
$m_{ij}^{(n)}$, $\varphi_i^{(n)}$ and $\psi_j^{(n)}$ satisfy the
differential and difference relations (\ref{relm}) and
(\ref{relphipsi}).

Next we impose the complex conjugate condition (\ref{cc}) with the
restriction (\ref{relx}). For this purpose, we consider general
rational solutions (\ref{e:taunK}) with $2N\times 2N$ determinants
(i.e., $K=2N$),
\begin{equation}
\tau_n=\det_{1\le i,j\le 2N}\left(m_{ij}^{(n)}\right),
\end{equation}
together with (\ref{mphipsi}) and (\ref{AB}). In this solution, we
impose the parameter conditions
\[
p_{N+i}=\frac{\epsilon}{\bar p_i},\quad
q_{N+j}=\frac{\epsilon}{\bar q_j},\quad
n_{N+i}=n_i,\quad
m_{N+j}=m_j,
\]
\[
c_{N+i,k}=\sum_{\mu=0}^k(-1)^{\mu}{n_i-\mu \choose k-\mu}\bar c_{i\mu},
\quad
d_{N+j,l}=\sum_{\nu=0}^l(-1)^{\nu}{m_j-\nu \choose l-\nu}\bar d_{j\nu},
\]
for $1\le i, j \le N$. Under these parameter conditions, we have
\[
\xi_{N+i}=\bar\xi_i,\quad
\eta_{N+j}=\bar\eta_j,\]
\[
A_{N+i}=\sum_{k=0}^{n_i}c_{N+i,k}(-\bar p_i\partial_{\bar p_i})^{n_i-k}
=(-1)^{n_i}\sum_{\mu=0}^{n_i}\bar c_{i\mu}
(\bar p_i\partial_{\bar p_i}-1)^{n_i-\mu},
\]
\[
B_{N+j}=\sum_{l=0}^{m_j}d_{N+j,l}(-\bar q_j\partial_{\bar q_j})^{m_j-l}
=(-1)^{m_j}\sum_{\nu=0}^{m_j}\bar d_{j\nu}
(\bar q_j\partial_{\bar q_j}-1)^{m_j-\nu}.
\]
Using the operator identities
\[ (\bar p_j\partial_{\bar p_j}-1)^k \bar p_j = \bar p_j (\bar p_j\partial_{\bar
p_j})^k, \qquad (\bar q_j\partial_{\bar q_j}-1)^k \bar q_j = \bar q_j (\bar q_j\partial_{\bar
q_j})^k,
\]
the elements of the determinant in $\tau_n$ become
\begin{eqnarray*}
&& \hspace{-2.5cm}   m_{i,N+j}^{(n)}=A_iB_{N+j}
\frac{\bar q_j}{p_i\bar q_j+\epsilon}
(-\epsilon p_i\bar q_j)^ne^{\xi_i+\bar\eta_j}
=(-1)^{m_j}\bar q_j A_i\bar B_j\frac{1}{p_i\bar q_j+\epsilon}
(-\epsilon p_i\bar q_j)^ne^{\xi_i+\bar\eta_j},
\\
&& \hspace{-2.5cm}  m_{N+i,j}^{(n)}=A_{N+i}B_j
\frac{\bar p_i}{\bar p_iq_j+\epsilon}
(-\epsilon\bar p_iq_j)^{-n}e^{\bar\xi_i+\eta_j}
=(-1)^{n_i}\bar p_i\bar A_iB_j
\frac{1}{\bar p_iq_j+\epsilon}
(-\epsilon\bar p_iq_j)^{-n}e^{\bar\xi_i+\eta_j},
\\
&& \hspace{-2.5cm}  m_{N+i,N+j}^{(n)}=A_{N+i}B_{N+j}
\frac{\epsilon\bar p_i\bar q_j}{\bar p_i+\bar q_j}
\Big(-\frac{\bar p_i}{\bar q_j}\Big)^{-n}e^{\bar\xi_i+\bar\eta_j}   \\ && \hspace{-0.8cm}
=(-1)^{n_i+m_j}\epsilon\bar p_i\bar q_j\bar A_i\bar B_j
\frac{1}{\bar p_i+\bar q_j}
\Big(-\frac{\bar p_i}{\bar q_j}\Big)^{-n}e^{\bar\xi_i+\bar\eta_j}.
\end{eqnarray*}
Since the $\tau_n$ solution can be scaled by an arbitrary constant,
we define a scaled $\tau_n$ function as
\[
\tau_n/\prod_{i=1}^N(-1)^{n_i+m_i}\epsilon\bar p_i\bar q_i  \to \tau_n.
\]
This scaled $\tau_n$ solution can be written as
\begin{equation} \label{e:gformula}
\tau_n=\left|\begin{matrix} m_{ij}^{(n)}
&\displaystyle\frac{(-1)^{m_j}}{\bar q_j}m_{i,N+j}^{(n)} \cr
\noalign{\vskip5pt} \displaystyle\frac{(-1)^{n_i}}{\epsilon\bar
p_i}m_{N+i,j}^{(n)} &\displaystyle\frac{(-1)^{n_i+m_j}}{\epsilon\bar
p_i\bar q_j}m_{N+i,N+j}^{(n)}\end{matrix} \right|
=\left|\begin{matrix} m_{ij}^{(n)} &\widehat{m}_{ij}^{(n)} \cr
\noalign{\vskip5pt}
\displaystyle\epsilon\overline{\widehat{m}_{ij}^{(-n)}}
&\displaystyle\overline{m_{ij}^{(-n)}}\end{matrix} \right|,
\end{equation}
where
\begin{equation} \label{e:mijhat}
\widehat{m}_{ij}^{(n)} \equiv
\displaystyle{\frac{(-1)^{m_j}}{\bar q_j}m_{i,N+j}^{(n)}}=A_i\bar
B_j\frac{1}{p_i\bar q_j+\epsilon} (-\epsilon p_i\bar
q_j)^ne^{\xi_i+\bar\eta_j}.
\end{equation}
We can see from (\ref{e:gformula}) that this $\tau_n$ satisfies the
complex conjugate condition (\ref{cc}), and thus it satisfies the
bilinear equation  (\ref{bilinDSII}) of the DSII equation.

Finally we simplify the above $\tau_n$ solution. Using the operator
identities
\[(p_i\partial_{p_i})p_i^ne^{\xi_i}=p_i^ne^{\xi_i} (p_i\partial_{p_i}+\xi'_i+n), \]
\[(q_j\partial_{q_j})(-q_j)^{-n}e^{\eta_j}=(-q_j)^{-n}e^{\eta_j} (q_j\partial_{q_j}+\eta'_j-n), \]
where
\begin{equation*}
\hspace{-2cm}
\xi_i'=-\frac{2}{p_i^2}x_{-2}-\frac{1}{p_i}x_{-1}+p_ix_1+2p_i^2x_2,\quad
\eta_j'=\frac{2}{q_j^2}x_{-2}-\frac{1}{q_j}x_{-1}+q_jx_1-2q_j^2x_2,
\end{equation*}
the rational solutions to the DSII equation can be obtained from
(\ref{AB}), (\ref{e:gformula}) and (\ref{e:mijhat}) as
\begin{equation} \label{e:final}
\tau_n=\left|\begin{matrix} m_{ij}^{(n)} &\widehat{m}_{ij}^{(n)} \cr
\noalign{\vskip5pt}
\displaystyle\epsilon\overline{\widehat{m}_{ij}^{(-n)}}
&\displaystyle\overline{m_{ij}^{(-n)}}\end{matrix} \right|,
\end{equation}
where
\begin{eqnarray*}
\hspace{-1.7cm}
m_{ij}^{(n)}=(-\frac{p_i}{q_j})^ne^{\xi_i+\eta_j}
\sum_{k=0}^{n_i}c_{ik}(p_i\partial_{p_i}+\xi'_i+n)^{n_i-k}
\sum_{l=0}^{m_j}d_{jl}(q_j\partial_{q_j}+\eta'_j-n)^{m_j-l}
\frac{1}{p_i+q_j},
\end{eqnarray*}
\begin{eqnarray*}
\hspace{-1.7cm}
\widehat{m}_{ij}^{(n)}=(-\epsilon p_i\bar
q_j)^ne^{\xi_i+\bar\eta_j}
\sum_{k=0}^{n_i}c_{ik}(p_i\partial_{p_i}+\xi'_i+n)^{n_i-k}
\sum_{l=0}^{m_j}\bar d_{jl}(\bar q_j\partial_{\bar q_j}+\overline{\eta'_j}+n)^{m_j-l}
\frac{1}{p_i\bar q_j+\epsilon}.
\end{eqnarray*}
Then using the gauge invariance of $\tau_n$, we see that $\tau_n$
with matrix elements
\begin{eqnarray*}
m_{ij}^{(n)}=
\sum_{k=0}^{n_i}c_{ik}(p_i\partial_{p_i}+\xi'_i+n)^{n_i-k}
\sum_{l=0}^{m_j}d_{jl}(q_j\partial_{q_j}+\eta'_j-n)^{m_j-l}
\frac{1}{p_i+q_j},
\end{eqnarray*}
\begin{eqnarray*}
\widehat{m}_{ij}^{(n)}=
\sum_{k=0}^{n_i}c_{ik}(p_i\partial_{p_i}+\xi'_i+n)^{n_i-k}
\sum_{l=0}^{m_j}\bar d_{jl}(\bar q_j\partial_{\bar q_j}+\overline{\eta'_j}+n)^{m_j-l}
\frac{1}{p_i\bar q_j+\epsilon},
\end{eqnarray*}
also satisfies the bilinear equation (\ref{bilin}) as well as the
complex conjugate condition (\ref{cc}), thus it satisfies the
bilinear equation (\ref{bilinDSII}) of the DSII equation. This
completes the proof of Theorem 1.

\section*{Appendix B}

In this appendix, we prove that $f$ in Theorem 1 is non-negative for
$\epsilon=-1$. In view of Eqs. (\ref{formula_rational}) and
(\ref{e:finaltau}), it suffices to show the following lemma.

\vspace{0.2cm} \textbf{Lemma 1.} \ For any $N\times N$ matrices $A$
and $B$, the following $2N\times 2N$ determinant is non-negative:
\[
\left|\begin{matrix} A &B \cr -\bar B &\bar A
\end{matrix} \right|\ge0.
\]

\vspace{0.2cm} \textbf{Proof.} \ We will prove this lemma by
induction. For $N=1$ the statement in the lemma is obviously true.
Let us denote $N\times M$ matrices as
\[
A_{NM}=\mathop{\rm mat}_{1\le i\le N,\,1\le j\le M}\Big(a_{ij}\Big),\quad
B_{NM}=\mathop{\rm mat}_{1\le i\le N,\,1\le j\le M}\Big(b_{ij}\Big),
\]
where $a_{ij}$ and $b_{ij}$ are complex numbers. Then by the Jacobi
formula for determinants, we have
\begin{eqnarray}  \label{e:Jacobi}
&&\left| \begin{matrix}A_{N+1,N+1} &B_{N+1,N+1} \cr
\noalign{\vskip5pt} -\overline{B_{N+1,N+1}}
&\overline{A_{N+1,N+1}}\end{matrix} \right| \left|
\begin{matrix}A_{NN} &B_{NN} \cr \noalign{\vskip5pt}
-\overline{B_{NN}} &\overline{A_{NN}}\end{matrix}\right| \cr
&&=\left| \begin{matrix}A_{N+1,N+1} &B_{N+1,N} \cr
\noalign{\vskip5pt} -\overline{B_{N,N+1}}
&\overline{A_{NN}}\end{matrix} \right| \left| \begin{matrix} A_{NN}
&B_{N,N+1} \cr \noalign{\vskip5pt} -\overline{B_{N+1,N}}
&\overline{A_{N+1,N+1}}\end{matrix} \right| \cr &&\quad -\left|
\begin{matrix}A_{N+1,N} &B_{N+1,N+1} \cr \noalign{\vskip5pt}
-\overline{B_{NN}} &\overline{A_{N,N+1}}\end{matrix} \right|
\left|\begin{matrix}A_{N,N+1} &B_{NN} \cr \noalign{\vskip5pt}
-\overline{B_{N+1,N+1}} &\overline{A_{N+1,N}}\end{matrix}\right|.
\end{eqnarray}
The right-hand side of this equation can be rewritten as
\[
\left|\begin{matrix} A_{N+1,N+1} &B_{N+1,N} \cr \noalign{\vskip5pt}
-\overline{B_{N,N+1}} &\overline{A_{NN}}\end{matrix} \right|^2
+\left| \begin{matrix}A_{N+1,N} &B_{N+1,N+1} \cr \noalign{\vskip5pt}
-\overline{B_{NN}} &\overline{A_{N,N+1}}\end{matrix} \right|^2,
\]
which is non-negative. Denoting
\[
D_N=\left|\begin{matrix} A_{NN} &B_{NN} \cr \noalign{\vskip5pt}
-\overline{B_{NN}} &\overline{A_{NN}}\end{matrix}\right|,
\]
then Eq. (\ref{e:Jacobi}) gives $D_{N+1}D_N\ge0$. Therefore if
$D_N>0$, we get $D_{N+1}\ge0$. If $D_N=0$, then by an infinitesimal
deformation (for example, $A_{NN}\to A_{NN}+\alpha I_N$ with an
infinitesimal real number $\alpha$ and the $N\times N$ unit matrix
$I_N$), the deformed $D_N$ becomes positive. Thus the
infinitesimally deformed $D_{N+1}$ is non-negative, and so is
$D_{N+1}$. This completes the induction and Lemma 1 is proved.

\section*{Appendix C}

In this appendix, we comment on the nonsingularity of rational
solutions for the DSI equation given in \cite{OY_DSI}. The solution
in Theorem 1 of \cite{OY_DSI} is nonsingular if the real parts of
wave numbers $p_i$ ($1\le i\le N$) are all positive. Because if
${\rm Re}\,p_i>0$, then from the appendix in \cite{OY_DSI}, it is
easy to see that the denominator $f$ is given by the determinant of
a Hermite matrix whose element can be written as an integral,
\[
f=\det_{1\le i,j\le N}\Big(m_{ij}^{(0)}\Big),\quad
m_{ij}^{(0)}=\int_{-\infty}^{x_1}A_i\bar A_je^{\xi_i+\bar\xi_j}\,dx_1.
\]
Here the condition of ${\rm Re}\,p_i>0$ (for all $1\le i\le N$) is
used to guarantee that the antiderivative of $e^{\xi_i+\bar\xi_j}$
(with respect to $x_1$) vanishes at $x_1=-\infty$. Then for any
non-zero vector $\mbox{\boldmath $v$}=(v_1,v_2,\cdots,v_N)$ and
${}^t\bar{\mbox{\boldmath $v$}}$ being its complex transpose, we have
\[
\mbox{\boldmath $v$}\Big(m_{ij}^{(0)}\Big)_{i,j=1}^N
{}^t\bar{\mbox{\boldmath $v$}}
=\int_{-\infty}^{x_1}\Big|\sum_{i=1}^Nv_iA_ie^{\xi_i}\Big|^2\,dx_1>0.
\]
This shows that the Hermite matrix $\Big(m_{ij}^{(0)}\Big)$ is
positive definite, hence its determinant $f$ is positive, i.e.,
$f>0$.

When the real parts of wave numbers $p_i$ ($1\le i\le N$) are all
negative, by slightly modifying the above argument, we can show that
the rational solutions in the DSI equation are nonsingular as well.

We conjecture that the rational solutions in the DSI equation, as
given in \cite{OY_DSI}, are actually nonsingular for all wave
numbers $p_i$ ($1\le i\le N$).


\section*{References}
\begin{thebibliography}{10}
\bibitem{rogue_water} C. Kharif, E. Pelinovsky and A. Slunyaev, \emph{Rogue waves in the
ocean} (Springer, Berlin, 2009).

\bibitem{Rogue_nature1} D. R. Solli, C. Ropers, P. Koonath and B. Jalali,
``Optical rogue waves", Nature 450, 1054--1057 (2007).

\bibitem{Peregrine}
D. H. Peregrine, ``Water waves, nonlinear Schr\"odinger equations
and their solutions," J. Australian Math. Soc. B, 25, 16–-43 (1983).

\bibitem{Akhmediev_PRE}
N. Akhmediev, A. Ankiewicz, and J. M. Soto-Crespo, ``Rogue Waves and
Rational Solutions of the Nonlinear Schrödinger Equation," Phys.
Rev. E 80, 026601 (2009).

\bibitem{Rogue_higher_order}
P. Dubard, P. Gaillard, C. Klein, V.B. Matveev, ``On multi-rogue
wave solutions of the NLS equation and positon solutions of the KdV
equation", Eur. Phys. J. Special Topics 185, 247-–258 (2010).

\bibitem{Rogue_higher_order2}
P. Dubard, V.B. Matveev, ``Multi-rogue waves solutions to the
focusing NLS equation and the KP-I equation", Nat. Hazards. Earth.
Syst. Sci. 11, 667--672 (2011).

\bibitem{Rogue_Gaillard}
P. Gaillard, ``Families of quasi-rational solutions of the NLS
equation and multi-rogue waves", J. Phys. A: Math. Theor. 44, 435204
(2011).

\bibitem{Rogue_triplet}
A. Ankiewicz, D.J. Kedziora and N. Akhmediev, ``Rogue wave
triplets", Phys. Lett. A, 375, 2782--2785 (2011).

\bibitem{Rogue_circular}
D.J. Kedziora, A. Ankiewicz, and N. Akhmediev, ``Circular rogue wave
clusters", Phys. Rev. E 84, 056611 (2011).

\bibitem{Liu_qingping}
B. Guo, L. Ling, and Q.P. Liu, ``Nonlinear Schr\"odinger equation:
Generalized Darboux transformation and rogue wave solutions", Phys.
Rev. E 85, 026607 (2012).

\bibitem{OY}
Y. Ohta and J. Yang, ``General high-order rogue waves and their
dynamics in the nonlinear Schr\"odinger equation", Proc. Roy. Soc.
A. 468, 1716–-1740 (2012).

\bibitem{Akhmediev_1985}
N. Akhmediev, V.M. Eleonskii, and N.E. Kulagin, ``Generation of a
periodic sequence of picosecond pulses in an optical fiber: Exact
solutions", Sov. Phys. JETP 89, 1542-–1551. [In Russian.]

\bibitem{Akhmediev_1988}
N. Akhmediev, V.M. Eleonskii and N.E. Kulagin, ``Exact first-order
solutions of the nonlinear Sch\"odinger equation", Theor. Math.
Phys. 72, 809--818 (1988).

\bibitem{Its_1988}
A.R. Its, A.V. Rybin, and M.A. Salle, ``Exact Integration of
Nonlinear Schr\"odinger equation", Theor. Math. Phys. 74, 29-–45
(1988).

\bibitem{Ablowitz_homo}
M.J. Ablowitz and B.M Herbst, ``On homoclinic structure and
numerically induced chaos for the nonlinear Schr6dinger equation",
SIAM J. Appl. Math. 50, 339--351 (1990).

\bibitem{Rogue_homo}
N. Akhmediev, A. Ankiewicz and M. Taki, ``Waves that appear from
nowhere and disappear without a trace", Phys. Lett. A 373, 675--678
(2009).

\bibitem{Ablowitz_private} M.J. Ablowitz, private communication (2012).

\bibitem{Rogue_nature2}
B. Kibler, J. Fatome, C. Finot, G. Millot, F. Dias, G. Genty, N.
Akhmediev, and J.M. Dudley, ``The Peregrine soliton in nonlinear
fibre optics", Nature Physics, 6, 790--795 (2010).

\bibitem{NLS_rogue_water}
A. Chabchoub, N. Hoffmann, M. Onorato, A. Slunyaev, A. Sergeeva, E.
Pelinovsky, and N. Akhmediev, ``Observation of a hierarchy of up to
fifth-order rogue waves in a water tank", Phys. Rev. E 86, 056601
(2012).

\bibitem{rogue_Hirota}
A. Ankiewicz, J. M. Soto-Crespo, and N. Akhmediev, ``Rogue waves and
rational solutions of the Hirota equation", Phys. Rev. E, 81, 046602
(2010).

\bibitem{rogue_DNLS1}
S. Xu, J. He and L. Wang, ``The Darboux transformation of the
derivative nonlinear Schrödinger equation", J. Phys. A  44, 305203
(2011).

\bibitem{rogue_DNLS2}
B. Guo, L. Ling and Q.P. Liu, ``High-order solutions and generalized
Darboux transformations of derivative nonlinear Schr\"odinger
equations", Stud. Appl. Math. DOI: 10.1111/j.1467-9590.2012.00568.x
(2012).

\bibitem{OY_DSI} Y. Ohta and J. Yang, ``Rogue waves in the Davey-Stewartson I equation," Phys. Rev. E 86, 036604 (2012).

\bibitem{Benney_Roskes}
D. J. Benney and G. Roskes,  ``Wave instabilities", Stud. Appl.
Math. 48, 377--385 (1969).

\bibitem{Davey_Stewartson} A. Davey and K. Stewartson,
``On three-dimensional packets of surface waves", Proc. R. Soc.
London, A 338, 101--110 (1974).


\bibitem{Ablowitz_book}
M.J. Ablowitz and H. Segur, \emph{Solitons and the Inverse
Scattering Transform} (SIAM, Philadelphia, 1981).

\bibitem{SA} J. Satsuma and M. J. Ablowitz,
``Two-dimensional lumps in nonlinear dispersive systems," J. Math.
Phys. 20, 1496--1503 (1979).

\bibitem{Ablowitz_book2}
M.J. Ablowitz and P.A. Clarkson, \emph{Solitons, Nonlinear Evolution
Equations and Inverse Scattering} (Cambridge University Press,
Cambridge, 1991).

\bibitem{Tajiri_1999} M. Tajiri and T. Arai, ``Growing-and-decaying mode solution
to the Davey-Stewartson equation," Phys. Rev. E 60, 2297--2305
(1999).

\bibitem{Nakamura_1982}
A. Nakamura, ``Explode-decay mode lump solitons of a two-dimensional
nonlinear Schr\"odinger equation", Phys. Lett. A 88, 55--56 (1982).

\bibitem{Nakamura_1983} A. Nakamura
``Exact explode-decay soliton solutions of a coupled nonlinear
Schr\"odinger equation," J. Phys. Soc. Jpn. 52, 3713--3721 (1983).

\bibitem{Ablowitz_Segur_1979}
M.J. Ablowitz and H. Segur, ``On the evolution of packets of water
waves," J. Fluid Mech. 92, 691--715 (1979).

\bibitem{Ablowitz_collapse2}
M.J. Ablowitz, I. Bakirtas, and B. Ilan, ``Wave collapse in a class
of nonlocal nonlinear Schr\"odinger equations," Physica D 207,
230--253 (2005).

\end{thebibliography}
\end{document}